# Theory of the tangential G-band feature in the Raman spectra of metallic carbon nanotubes

**S.M. Bose and S. Gayen**

Department of Physics, Drexel University, Philadelphia, PA 19104, USA

and

**S.N. Behera**

Physics Enclave, HIG 23/1, Housing Board Phase - I, Chandrasekharpur, Bhubaneswar 751016, Orissa, India

**Abstract**

The tangential G-band in the Raman spectra of a metallic single-wall carbon nanotube shows two peaks: a higher frequency component having the Lorentzian shape and a lower-frequency component of lower intensity with a Breit-Wigner–Fano (BWF)-type lineshape. This interesting feature has been analyzed on the basis of phonon-plasmon coupling in a nanotube. It is shown that the low-lying optical plasmon corresponding to the tangential motion of the electrons on the nanotube surface can explain the observed features. In particular, this theory can explain occurrence of both the Lorentzian and BWF lineshapes in the G-band Raman spectra of metallic single-wall carbon nanotubes. Furthermore, the theory shows that the BWF peak moves to higher frequency, has a lower intensity and a lower half width at higher diameters of the nanotube. All these features are in agreement with experimental observations.

# I. INTRODUCTION

Ever since the discovery of carbon nanotubes in 1991 by Iijima [1] there has been extensive research trying to understand their basic properties. Early on it was found both theoretically and experimentally that nanotubes of zero helicity are mostly metallic whereas nanotubes of non-zero helicity are mainly semiconducting [2,3].

Nanotubes have very interesting electronic and vibrational properties that have been studied by various methods. For example, the various peaks in the electron energy loss spectra of nanotubes have been identified with excitation of plasmons [4,5]. Raman spectroscopy has been used to study the vibrational modes in a nanotube [6]. Three dominant features have been identified in the Raman spectra of a single-wall carbon nanotube: a radial breathing mode (RBM) in the frequency range of 100-200 cm$^{-1}$, a vibrational mode (D line) around 1300 cm$^{-1}$ related to disordered carbons, and the G band in the frequency range 1580-1590 cm$^{-1}$ originating from the tangential oscillations of the carbon atoms in the nanotube. While in a semiconducting nanotube the G band shows one Raman peak as expected, in a metallic single-walled carbon nanotube (SWCNT) an additional band close to the main G band has been observed at somewhat lower frequencies (~1550 cm$^{-1}$). While the lineshape of the main G band is given by a Lorentzian, the additional lower frequency band can be fitted by a Breit-Wigner-Fano (BWF) lineshape [7,8]. It has been found experimentally that the peak of this side band moves to higher frequencies and that its height and width decreases with the increase of the radius of the SWCNT [9, 10]. The appearance of the Breit-Wigner-Fano line in carbon nanotube bundles has also been studied by Jiang et al. [11] who observed strong enhancement of this line because of bundling of the nanotubes.

The appearance of the BWF like side band and its dependence on the nanotube diameter has previously been investigated theoretically by several investigators [12,11,9]. All these authors have shown that the appearance of the extra peak only in metallic SWCNT is associated with the excitation of a plasmon in the nanotube and its interaction with the phonon responsible for the Raman spectra. These theories have explained the phenomenon in terms of an acoustic or semi-acoustic plasmon excited in the nanotube in the presence of momentum dependent defects. In the present paper we show that the observed phenomenon can also be explained properly if one invokes excitation of a low-lying optical plasmon corresponding to the azimuthal motion of the electrons on the surface of the metallic nanotube. Such a plasmon interacting with the phonon associated with tangential vibration of the carbon atoms gives rise to both the main Raman line as well as the lower frequency side band. The peak of the side band (BWF) is seen to move to higher frequency, and its height and the width decrease with the increase of the diameter of the nanotube, exactly what has been observed experimentally.

In section II we present the formulation of the present theory and in section III we present our results, compare them with experiments, and present a brief discussion.

## II. THEORY

During the Raman scattering process, light incident on a medium can be scattered with a lower frequency because of the emission of a phonon (Stokes lines), or it can be scattered with a higher frequency because of absorption of a phonon (anti-Stokes lines). At the microscopic level the incident electromagnetic wave first polarizes the medium by producing an electron-hole pair. The excited electron or the hole, in their turn, can either emit or absorb a phonon by the electron-phonon interaction process before recombining to produce the scattered photon. The intensity of the scattered photon (Raman line) is given by the spectral density of the phonon propagator renormalized by the electron-phonon interaction. This propagator must be calculated for a vanishingly small value of the wave vector q corresponding to the small value of q of the incident photon so that the momentum conservation is satisfied. Thus the electron polarization propagator responsible for the renormalization of the phonon propagator must also be calculated for small values of q. In an isotropic three-dimensional normal metal the electron polarization propagator vanishes for small values of q and hence there is no shift in the Raman lines due to electron-phonon coupling. The situation is quite different in a metallic carbon nanotube where, because of its cylindrical shape, the electron polarization propagator not only depends upon the wave vector q corresponding to electron motion along its axis but also on the quantum number $\mu$ corresponding to their azimuthal motion. It will be shown below that this polarization propagator has a pseudo-acoustic branch corresponding to $\mu=0$, which vanishes as q tends to zero. Thus this pseudo-acoustic branch of the polarization propagator should not be responsible for shift or split of the Raman line and hence would normally not be able to explain the appearance of BWF side-band in the G band of the Raman spectra of a metallic SWCNT, unless one introduces momentum dependent defects which are present in a nanotube. It turns out that the polarization propagator is non-zero for any integer value of $\mu$ ($\neq 0$) even when q=0, which is a consequence of the cylindrical shape of the nanotube. Therefore we expect a modification of the Raman line of a carbon nanotube for these values of the polarization propagator. As we shall see below this effect not only shifts the Raman line but produces a satellite band which can be identified with the BWF side band of the Raman G-band of a SWCNT. The positions and relative strengths of these two bands will depend on the plasmon-phonon interaction strength, the phonon and the plasmon frequencies and the diameters of the nanotubes.

As has been mentioned above, the Raman scattering intensity due to a phonon of frequency $\omega_q$ is related to the spectral density function (or the imaginary part) of the phonon propagator $D_Q(\omega)$ in the limit q→0 and hence can be written as [12]

$$I(\omega) \propto \mathrm{Im} D_Q(\omega)\big|_{q=0} ,\tag{1}$$

where the phonon propagator of the SWCNT, $D_Q(\omega)$, renormalized by the electron plasmon interaction can be expressed as

$$D_Q(\omega) = \frac{1}{[D_q^0(\omega)]^{-1} - (2\pi g)^2 \chi_Q(\omega)} .\tag{2}$$

In this equation the wave vector $Q=[q,\mu/a]$ is associated with the motion of the electrons in the nanotube, where q is the component of the wave vector in the direction of the nanotube axis and $\mu$ is the quantum number associated with the azimuthal motion of the electrons on the surface of the nanotube of radius a. The first term in the denominator of Eq. (2) is the inverse of the bare Green's function $D_q^0(\omega)$ of the nanotube phonon and is given by

$$D_q^0(\omega) = \frac{\omega_q}{\pi(\omega^2 - \omega_q^2)} .\tag{3}$$

The second term in the denominator of Eq. (2) is the contribution of the electron-phonon interaction. It has been expressed in terms of the electric susceptibility $\chi_Q(\omega)$ due to electron polarization of the nanotube and the electron-phonon interaction strength g. In the random phase approximation (RPA), the real part of the susceptibility function can be expressed as

$$\chi_Q(\omega) = \frac{\Pi_Q(\omega)}{1 - V(Q)\Pi_Q(\omega)} .\tag{4}$$

In this expression $\Pi_Q(\omega)$ is the electron polarization propagator and $V(Q)$ is the bare Coulomb interaction between electrons on the surface of the nanotube. In a previous publication dealing with the calculation of the plasmon frequencies in a metallic nanotube, we have calculated $\Pi_Q(\omega)$ for a metallic nanotube in the RPA and have shown it to be [14]

$$\mathrm{Re}\,\Pi_Q(\omega) \approx -\frac{k_F^2}{2\pi m \omega^2}(q^2 + \frac{\mu^2}{a^2})\tag{5}$$

where *m* is the mass of the electron in the nanotube and as has been mentioned before *a* is the radius of the nanotube. The imaginary part of $\Pi_Q(\omega)$ in the frequency range of plasmon excitation is vanishingly small and will be given an infinitesimal value of ε. In reference 14, the bare Coulomb interaction *V(Q)* has been shown to be

$$V(Q) = 4\pi e^2 a I_\mu(aq) K_\mu(aq) \qquad (6)$$

where *I* and *K* are the modified Bessel functions of appropriate arguments. Equations (5) and (6) show the effects of the cylindrical nature of the nanotube. For a metal of cubic symmetry the $\Pi_Q(\omega)$ and hence $\chi_Q(\omega)$ would tend to zero for small values of *q* and hence phonon propagator (Eq.2) and the Raman line would not get modified by the electron-phonon interaction. This will also be the case if we consider only the quasi-acoustic plasmon mode corresponding to $\mu=0$ in Eq. (5). In this case also both $\Pi_Q(\omega)$ and $\chi_Q(\omega)$ would tend to zero as *q* approaches zero and hence the Raman intensity will not be modified by the electron-phonon interaction, without the introduction of momentum defects. However, as can be seen from Eqs. (4) and (5), for $\mu\neq0$, $\Pi_Q(\omega)$ and $\chi_Q(\omega)$ are nonzero even when $q=0$ and hence the Raman line would be modified by the electron-phonon interaction as shown below.

Substituting Eqs. (5) and (6) in Eq. (4) and taking the limit $q\rightarrow 0$, we find that the dimensionless electronic susceptibility of the SWCNT takes the form

$$\tilde{\chi}_Q(\omega)_{q=0} \equiv N(0)\chi_Q(\omega)_{q=0} = \frac{(\omega_p^{0\,2}\mu^2 a_B/4a - i\varepsilon\omega^2)}{(\omega^2 - \omega_p^{0\,2}/4 + i\frac{a}{a_B}\varepsilon\omega^2)}, \qquad (7)$$

where $N(0)=m/2\pi$ is the density of states at the Fermi surface of the two-dimensional electron gas, $a_B$ is the Bohr radius, and $\omega_p^{0\,2} = \frac{4\pi n e^2}{ma}$ is associated with a classical plasma frequency. In Eq. (7) and below the symbol tilde on top of any character would indicate that the quantity has been expressed in dimensionless units.

Substituting Eqs. (7) and (3) in Eq. (2), we get the renormalized phonon propagator $D_Q(\omega)$ which when substituted in Eq. (1) would give us the Raman intensity as

$$I(\tilde{\omega}) = -\mathrm{Im}\,\pi\Omega_0 D_{q=0}(\omega+i\eta) = \frac{P(\tilde{\omega}^2 - \tilde{\omega}_p^2/4) - R\tilde{\varepsilon}\tilde{\omega}^2 a/a_0}{R^2 + P^2}, \qquad (8)$$

where

$$P = [2\tilde{\eta}\tilde{\omega}(\tilde{\omega}^2 - \mu^2\tilde{\omega}_p^{0\,2}/4) + \tilde{\varepsilon}(\tilde{\omega}^2 - 1 - \tilde{\eta}^2)\tilde{\omega}^2(a/a_B) + \lambda\varepsilon\omega^2],$$

and

$$R = (\tilde{\omega}^2 - 1 - \tilde{\eta}^2)(\tilde{\omega}^2 - \mu^2\tilde{\omega}_p^{0\,2}/4) - \tilde{\lambda}(a/a_B)\tilde{\omega}_p^{0\,2}/4 - 2\tilde{\eta}\tilde{\varepsilon}\tilde{\omega}^3 a/a_0,$$

where $\tilde{\lambda} = (\dfrac{\pi N(0)g^2}{\Omega_0})$ is the dimensionless parameter determining the strength of the electron phonon interaction. All parameters in this equation have been expressed in dimensionless form by normalizing them with respect to $\Omega_0 (\equiv \omega_q)$, the frequency of the phonon responsible for the G band in the Raman spectra of the SWCNT. It should be mentioned that in Eq. (8) we have added an infinitesimal width $\eta$ to the bare phonon by replacing $\omega$ with $\omega + i\eta$.

A careful examination of Eq. (8) will reveal that the Raman intensity will have two peaks near the renormalized frequencies of *1* and $\mu\tilde{\omega}_p^0/2$, corresponding to the excitation of the nanotube phonon and the optical plasmon with azimuthal quantum number $\mu$, respectively. As it will be seen in the next section, the first peak will correspond to the main G band whereas the second peak will correspond to the lower energy side band observed in the Raman spectra of SWCNT. It should be noted that for $\mu=0$ corresponding to excitation of a pseudo-acoustic plasmon, the second peak near $\mu\tilde{\omega}_p^0/2$ will drop out indicating that the pseudo-acoustic plasmon mode will not give rise to the side band. In order to explain the presence of the side bands in terms of pseudo-acoustic plasmons, previous authors [9,11,12] have introduced some extra momentum related to defects in the system. Our theory as described above explains the observed results without taking recourse to the presence of such defects.

## III. RESULTS AND DISCUSSION

The Raman intensity for the metallic SWCNT as given by Eq. (8) obviously will depend on the parameters $\lambda$, $\omega_p^{02}$, $a/a_B$, $\varepsilon$ and $\eta$. In Fig. 1, we have plotted the Raman intensity as a function of the frequency for three values of the radius $a/a_B = $ *12.5, 14.28,* and *16.66*, which correspond to the values of nanotube radius with which experiments have been performed [9]. In this figure we have chosen the effective electron-phonon interaction strength $\lambda$, the plasmon frequency $\widetilde{\omega}_p^{02}$, the infinitesimal width $\eta$ of the phonon frequency and the infinitesimal imaginary part ε of the polarization propagator to be *0.025, 3.80* and *0.02* and *.001*, respectively. In this calculation, we have chosen $\mu=1$, since this is the lowest possible frequency of an optical plasmon and easiest to excite.

As we can see, for each value of the radius, the Raman intensity has the main band along with a lower energy side band, which occurs because of the phonon-plasmon interaction during the Raman process. Indeed, the main Raman band seems to have a Lorentzian shape and the side band is more asymmetric and looks like the BWF peak observed in SWCNT [9,10]. If we now examine the peak locations of the side band, we notice that they move to higher frequencies with the increase of the radius of the nanotube. This diameter dependence of the peak frequency of the side band is exactly what has been observed experimentally [9,10]. Also an examination of the heights and widths of the peaks of the side bands in Fig. 1 indicates that they decrease as the diameter of the SWCNT increases. Although the experimental data do not seem to address the issue of diameter dependence of the height, experimental results clearly show that width of the BWF band decreases with the increase in the diameter of the SWCNT [9], in agreement with our calculation. Changing the values of the parameters $\lambda$, $\omega_p^{02}$, $a/a_B$, $\varepsilon$ and $\eta$ to other nonzero values give us results which are similar to those shown in Fig. 1.

To study more fully the diameter dependence of the width of the side band, we have plotted in Fig. 2, the Raman intensity for the side band with the same values of the parameters except that we have set $\varepsilon=0$. In the absence of the width of the plasmon propagator ($\varepsilon=0$), the main band (not shown in Fig. 2) and the side band get fully separated, which allows us to evaluate the width of the side band. A careful examination of the full width at half maximum (FWHM) of the side band of the three curves shows that the FWHM is smaller for the nanotubes with higher radii, which agree with experiments. As seen in Fig. 1, our theory also gives a shift of the peak position of the main band to lower frequencies with the increase of the nanometer diameter, although by a somewhat lower amount. The higher frequency peak in the experimental spectra shows only a weak dependence on the SWCNT diameter. As far as we can ascertain, previous theories depending on the excitation of pseudo-acoustic plasmon, do not address the issues of diameter dependence of the location, height and width of the BWF side band, or the amount of shift in the location of the peak in the main band.

In conclusion, in this paper we have developed a theory of the G band Raman spectra of a SWCNT based on the simultaneous excitation of an optical plasmon associated with azimuthal motion of the electrons on the nanotube surface in a metallic nanotube. Our theory shows that this low-lying optical plasmon can interact with the nanotube phonon giving rise to the BWF line along with the main G band in the Raman spectra. The various features of the side band calculated in this paper agree with experimental observations.

**ACKNOWLEDGEMENT**

SMB and SNB would like to acknowledge the hospitality of the School of Physics, University of Hyderabad, India, where the manuscript was prepared.

# FIGURE CAPTIONS

1. (Color online) Raman intensity of metallic carbon nanotubes as obtained from Eq. (8) for three values of the nanotube radius and with values of the parameters given in the text. The dot-dashed, solid, and dashed curves are obtained for $a/a_B$ = 12.5, 14.28 and 16.66, respectively. Note the diameter dependence of the frequency, width and height of the peak of the side band.

2. (Color online) Raman spectra of metallic carbon nanotubes with the same parameters of Fig.1 except that here $\varepsilon$ have been set to be 0. This figure allows careful examination of the diameter dependence of the FWHM of the side band.

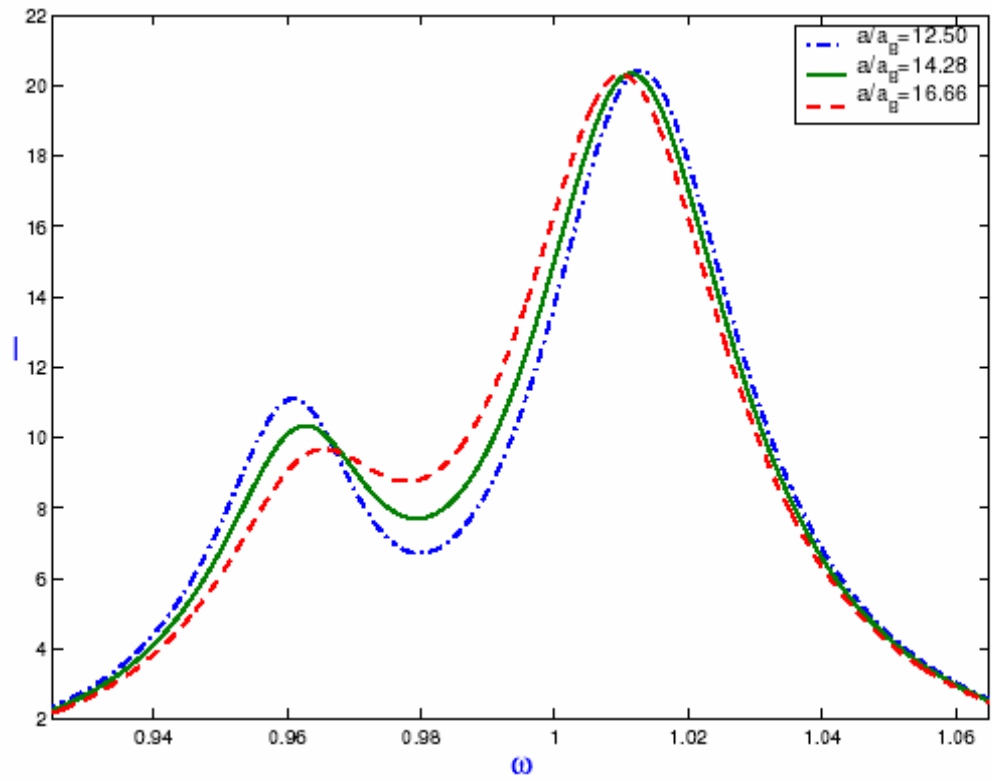

Figure 1

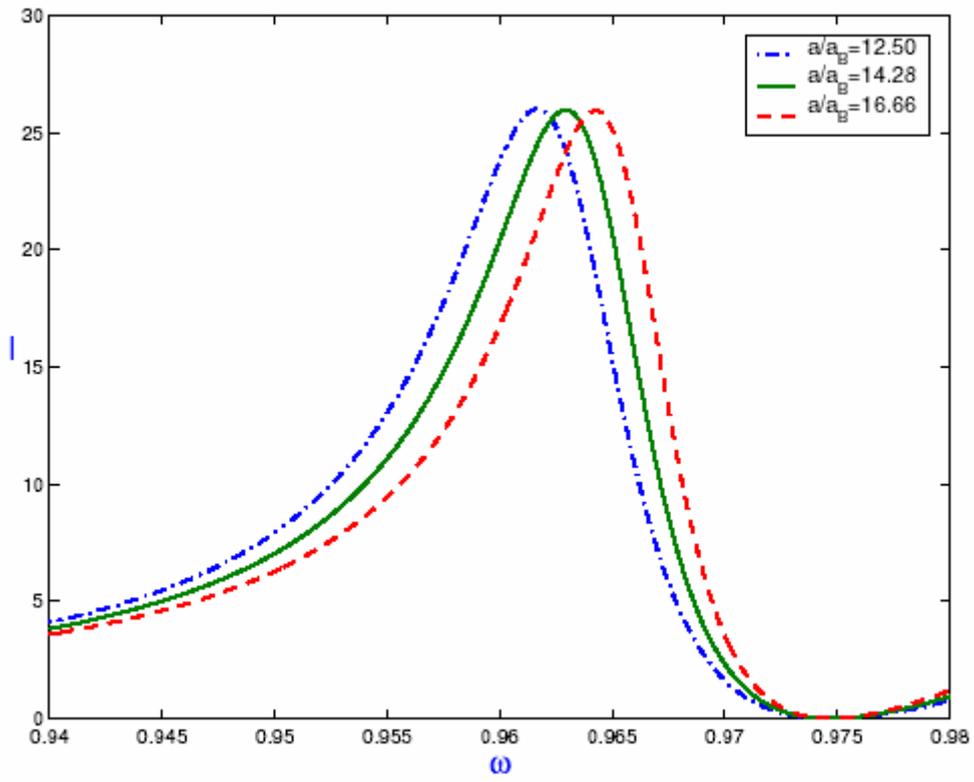

Figure 2